\documentclass[prd,nofootinbib,nobibnotes,showpacs,superscriptaddress]
              {revtex4}
\usepackage{graphicx}
\usepackage{amssymb}

\begin{document}

\title{Faraday Rotation of the Cosmic Microwave Background Polarization
       by a Stochastic Magnetic Field}

\author{Arthur Kosowsky}
\email{kosowsky@physics.rutgers.edu}
\affiliation{Department of Physics and Astronomy, Rutgers University,
136 Frelinghuysen Rd, Piscataway, NJ 08854}
\author{Tina Kahniashvili}
\email{tinatin@phys.ksu.edu}
\affiliation{Department of Physics, Kansas State University,
116 Cardwell Hall, Manhattan, KS 66506, and \\
Center for Plasma Astrophysics, Abastumani Astrophysical Observatory,
2A Kazbegi Ave., GE-0160 Tbilisi, Georgia}
\author{George Lavrelashvili}
\email{lavrela@itp.unibe.ch}
\affiliation{Department of Theoretical Physics, A.~Razmadze Mathematical
Institute, GE-0193 Tbilisi, Georgia}
\author{Bharat Ratra}
\email{ratra@phys.ksu.edu}
\affiliation{Department of Physics, Kansas State University,
116 Cardwell Hall, Manhattan, KS 66506}

\date{September 2004 \hspace{0.3truecm} KSUPT-04/5}

\begin{abstract}
A primordial cosmological magnetic field induces Faraday rotation of the
cosmic microwave background polarization. This rotation produces a curl-type
polarization component even when the unrotated polarization possesses only
gradient-type polarization, as expected from scalar density perturbations.
We compute the angular power spectrum of curl-type polarization arising from
small Faraday rotation due to a weak stochastic primordial magnetic field
with a power-law power spectrum. The induced polarization power spectrum
peaks at arcminute angular scales. Faraday rotation is one of the few
cosmological sources of curl-type
polarization, along with primordial tensor perturbations,
gravitational lensing, and the vector and tensor perturbations induced
by magnetic fields; the Faraday rotation signal peaks on
significantly smaller angular scales than any of these, with a power spectrum
amplitude which can be comparable to that from gravitational lensing.
Prospects for detection are briefly discussed.
\end{abstract}


\maketitle

\section{Introduction}
\label{sec:intro}

A primordial cosmological seed magnetic field has been proposed to
explain the existence of observed large-scale ($\sim$ 10 kpc), $\mu$G
strength, ordered magnetic fields observed in galaxies and galaxy
clusters \cite{review,inflation}.  Cosmological perturbations induced
by such a magnetic field and the corresponding cosmic microwave
background temperature and polarization anisotropies have been the
subject of a number of recent studies \cite{chen95,adams96,durrer98,
jedamzik98,subramanian98a,subramanian98b,jedamzik00,durrer00,
koh00,seshadri01,mack02,pogosian02,
subramanian03,caprini03,lewis04}.  The strength of such a hypothetical
cosmological magnetic field can be significantly constrained by
comparing model predictions and microwave background observational
data \cite{WMAP}.

Several studies of the evolution of cosmological magnetic fields with
power-law power spectra have been done \cite{banerjee04,dolag99}.
The strongest constraints
result from amplification of primordial magnetic fields via an inverse
cascade mechanism, which takes power on small scales and transfers it
to larger scales; this process is effective for fields with significantly
increasing power on small scales \cite{boldyrev03,verma04}.
Such ``blue'' power spectra result from magnetic fields generated in small
scales at late time, for example, during cosmological
phase transitions. In this work we consider general power-law magnetic
fields, including fields generated during an early inflation-epoch, which are
not strongly constrained by their subsequent evolution. Observational
and theoretical constraints on primordial fields are discussed in more
detail in the concluding section of this paper.

A distinctive method of constraining a cosmological magnetic field is
to study the rotation of the microwave background polarization
orientation due to the Faraday effect, as the radiation propagates
to us in the cosmological magnetic field. A homogeneous magnetic field
of strength $10^{-9}$ gauss induces a measurable rotation of order
$1^\circ$ at a frequency of $30$ GHz \cite{kosowsky96}. Faraday rotation
is the classic method of probing large-scale magnetic fields in the
universe, using quasars as sources of polarized radiation, e.g.~\cite{zuk99}.
Faraday rotation of radiation from discrete point sources can constrain
the primordial magnetic field power spectrum \cite{kolatt98}, and can
also probe the evolution of the magnetic field with redshift \cite{sethi03}.
The microwave background in principle offers a great advantage because it
provides a source of polarized light from a fixed redshift which covers
the entire sky; with sufficiently precise measurements the projected
magnetic field can be mapped in detail. The drawback is that the
polarization of the microwave background is very small, only a part in a
million, and the Faraday rotation expected from a cosmological magnetic field
is also small and so challenging to detect.

In general, a polarization field has two independent components, known
in the microwave background context as $G$ and $C$ \cite{kks97b} (for
``gradient'' and ``curl''), or equivalently as $E$ and $B$ modes
\cite{zaldarriaga97}. It is well known that primordial scalar density
perturbations induce only $G$-polarization, and that $C$ polarization must
arise from other production processes, notably primordial tensor
perturbations \cite{kamionkowski97a,seljak97}, gravitational lensing of the
$G$-polarization component \cite{zaldarriaga98}, and vector and tensor
perturbations from, e.g., magnetic fields \cite{mack02}. Faraday rotation,
which rotates polarization orientations, can also be a source of
$C$-polarization given an intial $G$-polarization field.

In this paper we compute the power spectrum of microwave background
$C$ polarization resulting from Faraday rotation by a stochastic
primordial magnetic field with a given power spectrum and helicity
spectrum. We find that the power spectrum distinctively peaks on
arcminute angular scales, significantly smaller than other commonly
considered sources of $C$ polarization.  Previous work on the
polarization power spectrum has considered the simpler but less
realistic special case of a constant magnetic field
\cite{kosowsky96,scannapieco97,scoccola04}, or considered a stochastic
magnetic field but only calculated the power spectrum of the rotation
measure rather than of the microwave background polarization
\cite{campanelli04}. Sophisticated analysis techniques for extracting
microwave background power spectra from maps of the sky have been
developed, so measuring the power spectrum of Faraday rotation and
then confirming its scaling with frequency will be much more efficient
than requiring frequency information from the outset to measure the
rotation.  Measuring the power spectrum of Faraday rotation also
sidesteps any systematic errors associated with comparing observations
at different frequencies; this is a significant advantage since these
measurements will all be dominated by systematic errors.

The rest of the paper is organized as follows. The next Section
defines the primordial magnetic field power spectrum and helicity
spectrum and explains the approximation we employ to separate
polarization generation from polarization orientation rotation at the
surface of last scattering.  Section III derives the power spectrum of
the Faraday rotation measure, which distills the Faraday effect of the
tangled primordial magnetic field into a simple scalar function on the
sky. We give an explicit demonstration that any helical component of
the magnetic field does not contribute to Faraday rotation. Then given
this rotation field and a primordial $G$-polarization field, Sec.~IV
derives an analytic expression for resulting power spectrum of $C$
polarization. Section V numerically evaluates these expressions and
presents the power spectra for both the rotation field and the $C$
polarization for a range of magnetic field power spectra, and the
final Section discusses these results in the context of future
experiments to measure small-scale microwave background polarization.
Two short appendices contains some useful but technical mathematical
results.

\section{Magnetic Field Model}
\subsection{Magnetic Field Power Spectrum}

We will assume the existence of a cosmological magnetic field
generated during or prior to the early radiation-dominated epoch,
with the energy density of the magnetic field a first-order
perturbation to the standard Friedmann-Lema\^\i tre-Robertson-Walker
homogeneous cosmological spacetime model. The conductivity of the
primordial plasma is high so we can work in the
infinite-conductivity limit for all scales larger than the damping
scale set by photon and neutrino diffusion. This results in a
``frozen-in'' magnetic field and a corresponding electric field
${\bf E}=-{\bf v}\times {\bf B}$ where ${\bf v}$ is the plasma fluid
velocity. Since the fluid velocity is always small in the early
universe on the scales considered, the electric field is always
negligible compared to the magnetic field and will not be considered
further. Neglecting fluid back-reaction onto the magnetic field, the
spatial and temporal dependence of the magnetic field separates,
with magnetic flux conservation fixing the temporal dependence to be
the simple scaling ${\mathbf B}({\mathbf x}, \eta)={\mathbf
B}({\mathbf x})/a^2$ where $a$ is the scale factor and $\eta$
conformal time.

We also assume the magnetic field is a Gaussian random field.
Taking into account the possible helicity of the field \cite{vachaspati01},
the magnetic field spectrum in wavenumber space is
\cite{pogosian02,caprini03}
\begin{equation}
   \langle B^*_i({\mathbf k})B_j({\mathbf k'})\rangle
   =(2\pi)^3 \delta^{(3)}\!({\mathbf k}-{\mathbf k'}) [P_{ij} P_B(k)
   + i \epsilon_{ijl} \hat{k}_l P_H(k)],
   \label{spectrum}
\end{equation}
where $P_B(k)$ and $P_H(k)$ are the symmetric and helical parts of the
magnetic field power spectrum, $\epsilon_{ijl}$ is the antisymmetric
tensor, and the plane projector
$P_{ij}\equiv\delta_{ij}-\hat{k}_i\hat{k}_j$ satisfies
$P_{ij}P_{jk}=P_{ik}$ and $P_{ij}\hat{k}_j=0$ with unit wavenumber
components $\hat{k}_i=k_i/k$.  We use the convention
\begin{equation}
   B_j({\mathbf k}) = \int d^3\!x \,
   e^{i{\mathbf k}\cdot {\mathbf x}} B_j({\mathbf x}),
   \qquad
   B_j({\mathbf x}) = \int {d^3\!k \over (2\pi)^3}
   e^{-i{\mathbf k}\cdot {\mathbf x}} B_j({\mathbf k})
\end{equation}
when Fourier transforming between real and wavenumber spaces.

The power spectrum $P_B(k)$ is related to the energy density of the
magnetic field, while the helicity part $P_H(k)$ is related to the spatial
average $\langle {\mathbf B} \cdot (\nabla \times {\mathbf B}) \rangle$
\cite{pogosian02}. Transforming from an orthonormal basis
$\{{\bf e}_1, {\bf e}_2, {\bf e_3} = {\bf\hat k}\}$ to the helicity
basis \cite{varshalovich89}
\begin{equation}
   {\mathbf e}^{\pm }({\bf k})
   =-\frac{i}{\sqrt{2}}({\mathbf e}_1 \pm  i{\mathbf e}_2), \qquad
   {\bf e_3} = {\bf\hat k},
   \label{helicity-basis}
\end{equation}
the power spectra $P_B(k)$ and $P_H(k)$ can be expressed in terms of
magnetic field components in the helicity basis
as \cite{pogosian02,caprini03}
\begin{eqnarray}
   \langle B^{+}{(\mathbf k)} B^{+}(-{\mathbf k^\prime}) +
   B^{-}({\mathbf k}) B^{-}(-{\mathbf k^\prime}) \rangle &=& -(2\pi)^3 P_B(k)
   \delta^{(3)}\!({\mathbf k} - {\mathbf k^\prime})~,\label{PB-basis}\\
   \langle B^{+}({\mathbf k}) B^{+}(-{\mathbf k^\prime)} -
   B^{-}({\mathbf k}) B^{-}(-{\mathbf k^\prime})\rangle &=& (2\pi)^3 P_H(k)
   \delta^{(3)}\!({\mathbf k} - {\mathbf k^\prime}).
   \label{PH-basis}
\end{eqnarray}
We describe both the symmetric and helical parts
by simple power laws
\begin{equation}
   P_B(k)=A_B k^{n_B},\qquad P_H(k)=A_H k^{n_H}.
   \label{power-law}
\end{equation}
If the magnetic field is generated in small scales, after inflation,
the power spectrum index
is constrained to be $n_B \geq 2$ \cite{durrer03}. Also the
amplitudes are generically constrained by $P_B(k) \geq |P_H(k)|$
\cite{durrer03,verma04,banerjee04},
which implies $n_H \geq n_B$ \cite{caprini03,durrer03}:
a field cannot support a fixed helicity in the limit of zero field
strength. Note that some authors define spectral indices which
correspond to our $n_B+3$ and $n_H+3$
\cite{sigl97,banerjee04}.

As a phenomenological normalization of the magnetic field, we smooth
the field on a comoving scale $\lambda$ by convolving with a Gaussian
smoothing kernel $f_\lambda({\bf x}) = N\exp(-x^2/2\lambda^2)$
to obtain the smoothed field ${\bf B}_\lambda({\bf x})$,
and introduce average  values of energy density \cite{mack02}
\begin{equation}
   {B_\lambda}^2 \equiv \langle {\mathbf B}_\lambda({\mathbf x})
   \cdot {\mathbf B}_\lambda({\mathbf x})\rangle
   = {1\over \pi^2}\int_0^\infty dk \, k^2 \, e^{-k^2\lambda^2}P_B(k),
   \label{Blambda}
\end{equation}
and helicity \cite{caprini03}
\begin{equation}
   {H_\lambda}^2 \equiv
   \lambda|\langle {\mathbf B}_\lambda({\mathbf x}) \cdot
   \{{\mathbf \nabla} \times {\mathbf B}_\lambda ({\mathbf x})\}\rangle|
   = {\lambda\over \pi^2}\int_0^\infty dk \, k^3 \, e^{-k^2\lambda^2}|P_H(k)|.
   \label{Hlambda}
\end{equation}
Equations (\ref{Blambda}) and (\ref{Hlambda}) may be used to re-express
the power-law power spectra of Eqs.~(\ref{power-law}) in closed form as
\cite{mack02,pogosian02,caprini03}
\begin{eqnarray}
   P_B(k)=\frac{(2\pi)^{n_B+5}}{2}
   \frac{B_\lambda{}^2}{\Gamma\left(n_B/2 + 3/2\right)}
   \frac{k^{n_B}}{k^{n_B+3}_\lambda},\qquad k<k_D,
   \label{energy-spectrum-S} \\
   P_H(k)=\frac{(2\pi)^{n_H+5}}{2}\frac{H_\lambda{}^2}
   {\Gamma\left(n_H/2 + 2\right)}\frac{k^{n_H}}{k^{n_H+3}_\lambda},
   \qquad k<k_D,
   \label{energy-spectrum-H}
\end{eqnarray}
where the smoothing wavenumber $k_\lambda=2\pi/{\lambda}$. We make
the approximation that both spectra vanish for all wavenumbers $k$
larger than a damping wavenumber $k_D$. We assume that the magnetic field
damping  is due to Alfv\'en wave damping from photon viscosity,
and the cut-off wavenumber to be \cite{mack02,jedamzik98,subramanian98a}
\begin{equation}
  \left({k_D \over {\rm Mpc}^{-1}}\right)^{n_B + 5}
  \approx 2.9\times 10^4
  \left({B_\lambda\over 10^{-9}\,{\rm G}}\right)^{-2}
  \left({k_\lambda\over {\rm Mpc}^{-1}}\right)^{n_B + 3} h,
  \label{damping-scale}
\end{equation}
which  will always be a much smaller scale than the Silk damping scale
(thickness of the last scattering surface) for standard cosmological
models ($h$ is the Hubble constant in units of 100 km s$^{-1}$ Mpc$^{-1}$).

\subsection{Faraday Rotation}

A cosmological magnetic field at the epoch of last scattering will rotate
the CMB polarization orientation in a given sky direction due to Faraday
rotation (see, e.g., \cite{landau81}).  For a review of CMB polarization
theory see \cite{kosowsky99}; for computational methods and statistics, see
\cite{kosowsky96a,hu97,kks97b,zaldarriaga97}. The time derivative of the
orientation angle $\alpha$ of linearly polarized monochromatic radiation
passing through a plasma in the presence of a magnetic field
$\mathbf B({\mathbf x}, \eta)$ is \cite{kosowsky96}
\begin{equation}
   \omega_{\mathbf B}({\bf x}, {\mathbf n}, \eta)={\dot \alpha}
   \equiv \frac{d\alpha}{d\eta}=\frac{q^3 x_e n_e a}{2\pi m_e^2
   \nu^2} \, {\mathbf B}({\mathbf x}, \eta) \cdot {\mathbf n} .
   \label{omegaB}
\end{equation}
Here  ${\mathbf n}$ is the propagation direction of the radiation and
$\nu$ the radiation frequency. The electron charge and mass are $q$ and
$m_e$, and $n_e$ and $x_e$ are the total electron number density and
ionization fraction.  We normalize the scale factor $a$ by setting
$a_0 = 1$ today. We employ cosmological units with $\hbar = 1 = c$.

In terms of the comoving magnetic field ${\mathbf B}({\mathbf
x})={\bf B}({\mathbf x}, \eta)a^2$, Eq.~(\ref{omegaB}) can be
rewritten as
\begin{equation}
   \omega_B({\bf x}, {\bf n}, \eta)
   =\frac{3}{(4\pi)^2\nu_0^2 q} \, \dot\tau({\bf x}) \, {\bf B}({\bf x})
   \cdot {\bf n}
   \label{omegaB2}
\end{equation}
with comoving frequency of the observed radiation $\nu_0$ and
the differential optical depth ${\dot \tau}=x_e n_e \sigma_T a$.
Here $\sigma_T=8\pi\alpha_{EM}^2/(3m_e^2)$ is the Thompson scattering
cross section with the fine-structure constant $\alpha_{EM} = q^2 \approx
1/137$. We neglect inhomogeneities in the free
electron density, assuming that $\dot\tau(\eta)$ is independent of
${\bf x}$; then the rotation angle observed today is
\begin{equation}
   \alpha({\bf n},\eta_0) = \frac{3}{(4\pi)^2\nu_0^2 q}
   \int_{\eta_{dec}}^{\eta_0} d\eta \,
   \dot\tau(\eta) \, {\bf B}({\bf x}) \cdot {\bf n} .
   \label{phi}
\end{equation}
Throughout the rest of this paper, ${\bf B}$ represents the comoving
value of the magnetic field.

Faraday rotation of the microwave background is a somewhat subtle
problem, because the polarization is generated and rotated
simultaneously in the region of the last scattering surface. In a
rigorous treatment, these two effects must be computed together
via the complete radiative transfer equations describing the evolution
of polarization fluctuations. However, as long as the total rotation
is small compared to $\pi/2$, the total rotated polarization angle can
be expressed simply as an average of the rotated polarization angle from
each infinitesimal piece of path length through the surface of last
scattering, neglecting depolarization effects. We make
the simplifying approximation that any magnetic field component
with a wavelength shorter than the thickness of the surface of last
scattering is neglected. For these components, the rotation of
polarization generated at different optical depths will tend to
cancel, leaving little net rotation.  (This assumption breaks down if
the magnetic field strength is dominated by components on small
scales. In this case, a more realistic power-law dropoff
in $P_B(k)$ should be used instead of a sharp cutoff.) This is equivalent
to imposing a cutoff scale on the magnetic field at the photon damping
scale at last scattering. Then we can treat the magnetic field as
constant throughout the rotation region, so that the total rotation,
which is the sum over the rotations of each infinitesimal piece of
generated polarization, can be expressed as the total rotation
incurred by the polarization generated at some particular effective
optical depth, which we denote by $\eta_{\rm rot}$.

Assuming that $\eta_{\rm rot}$ is the conformal time corresponding to
$\tau = 1$ (the actual value is a little lower since the
polarization visibility function peaks at a lower redshift than the
temperature visibility function), we can derive a simple expression
for the approximate Faraday rotation in a given observation direction.
Writing ${\bf x} = {\bf n} (\eta_0-\eta)$ (i.e., putting the observer
at the origin of the coordinate system) and Fourier expanding the magnetic
field, Eq.~(\ref{phi}) can be written as
\begin{equation}
   \alpha({\bf n}, \eta_0) \simeq \frac{3}{4 (2 \pi)^5\nu_0^2 q}
   \int d^3\!k \, {\bf B}({\bf k}) \cdot {\bf n} \,
   e^{-i{\bf k}\cdot{\bf n}\,\Delta\eta},
   \label{phisol}
\end{equation}
where $\Delta\eta = \eta_0 - \eta_{\rm rot}$.  These approximations
have split the polarization generation and rotation problems; we can
treat the full problem as equivalent to the generation of polarization
in the usual way, followed by rotation using an effective rotation
screen just prior to the radiation reaching the observer. The rest of
this paper is devoted to computing the properties of the effective
rotation screen and the resulting power spectrum of microwave
background polarization.

\section{Rotation Power Spectrum}
\label{average_rm}

It is conventional to introduce a wavelength-independent measure of the
rotation of the polarization orientation, the rotation measure
$R({\bf n})\equiv\alpha({\bf n})\nu_0^2$. The two-point correlation
function of the effective rotation measure for a stochastic magnetic
field based on the approximate solution in Eq.~(\ref{phisol}) is
\begin{equation}
   \left\langle R({\bf n}) R({\bf n^\prime})\right\rangle
   \simeq \frac{9}{16 (2\pi)^{10} q^2} \int
   d^3\!k \int d^3\!k^\prime e^{i{\bf k}\cdot{\bf n}\,\Delta\eta}
   e^{-i{\bf k^\prime}\cdot{\bf n^\prime}\,\Delta\eta} n_i n_j^\prime
   \left\langle B^*_i({\bf k}) B_j({\bf k^\prime})\right\rangle
   \label{RR'0}
\end{equation}
where $\Delta\eta \equiv \eta_0-\eta_{\rm rot}\approx\eta_0$.

Prior to the ensemble averaging, it is convenient to decompose the
vector plane wave into vector spherical harmonics \cite{varshalovich89}
\begin{equation}
   \mathbf{B}(\mathbf{k}) e^{i{\mathbf{k \cdot n}}\eta_0}
   = \sum_{l= 0}^{\infty} \sum_{m = -l}^{l}\sum_{\lambda=-1}^1
   A_{lm}^{(\lambda)}({\bf k})
   \mathbf{Y}_{lm}^{(\lambda)}(\mathbf{n})
   \label{vdecom}
\end{equation}
where $\lambda$ labels the three orthonormal vector spherical harmonics
defined by
\begin{eqnarray}
   \mathbf{Y}_{lm}^{(1)}(\mathbf{n})&=&\frac{1}{\sqrt{l(l+1)}}
   \mathbf{\nabla}_{\bf n}
   Y_{lm}(\mathbf{n}), \nonumber\\
   \mathbf{Y}_{lm}^{(0)}(\mathbf{n})&=&\frac{-i}{\sqrt{l(l+1)}}
   [\mathbf{n} \times
   \mathbf{\nabla}_{\bf n}] Y_{lm}(\mathbf{n}), \nonumber\\
   \mathbf{Y}_{lm}^{(-1)}(\mathbf{n})&=&\mathbf{n}
   Y_{lm}(\mathbf{n}).
   \label{vectorspherical}
\end{eqnarray}
Here ${\bf\nabla}_{\bf n}$ is the two-dimensional covariant derivative
orthogonal to the unit vector ${\bf n}$, so the harmonics $\lambda=1$
and $\lambda=0$ are transverse to ${\bf n}$, while the $\lambda=-1$
harmonic is parallel to ${\bf n}$. When the expansion (\ref{vdecom})
is substituted into Eq.~(\ref{RR'0}), only the $\lambda=-1$ terms
contribute since the magnetic fields only appear through the contraction
${\bf n}\cdot{\bf B}$. Explicit forms for the coefficients
$A_{lm}^{(\lambda)}({\bf k})$ are given in Ref.\ \cite{varshalovich89};
the one needed here is
\begin{eqnarray}
A_{lm}^{(-1)} &=& {4\pi\over 2l+1} i^{l-1}
\left\{\sqrt{l(l+1)}\left[j_{l+1}(k\eta_0)+j_{l-1}(k\eta_0)\right]
{\bf B}({\bf k})\cdot{\bf Y}_{lm}^{(1)\,*}({\bf\hat k})\right.\nonumber\\
&&\qquad\qquad\qquad\qquad
- \left.\left[(l+1)j_{l+1}(k\eta_0)-lj_{l-1}(k\eta_0)\right]
{\bf B}({\bf k})\cdot{\bf Y}_{lm}^{(-1)\,*}({\bf\hat k})\right\}.
\end{eqnarray}
The second term proportional to ${\bf Y}_{lm}^{(-1)}$ is zero since
the magnetic field is divergenceless,
${\bf k}\cdot{\bf B}({\bf  k})=0$, leaving
\begin{equation}
   A_{lm}^{(-1)}=4\pi i^{l-1} \sqrt{l(l+1)}
   \frac{j_l(k\eta_0)}{k\eta_0} \mathbf{B}(\mathbf{k}) \cdot
   \mathbf{Y}_{lm}^{(1)\,*} (\mathbf{\hat k}) .
   \label{11}
\end{equation}
With these definitions the rotation angle becomes
\begin{equation}
   \alpha({\mathbf n}, \eta_0) \simeq \frac{3}{32\pi^4\nu_0^2 q}
   \int d^3\!k \sum_{lm}  i^{-l+1}
   \sqrt{l(l+1)} \frac{j_l(k\eta_0)}{k\eta_0} \, \mathbf{B}(\mathbf{k})
   \cdot \mathbf{Y}_{lm}^{(1) \star} (\mathbf{\hat k}) \,
   Y_{lm}(\mathbf{n}),
  \label{phisol2}
\end{equation}
and for the magnetic field spectrum of Eq.~(\ref{spectrum}) we obtain
\begin{eqnarray}
   \left\langle R({\mathbf n}) R({\mathbf n^\prime})\right\rangle
   &\simeq& \frac{9}{128\pi^5q^2} \int dk \, k^2
   P_B(k) \int d\Omega_{\mathbf{\hat k}} \nonumber\\
   && \times \sum_{lm}
   \sum_{l'm'} i^{l'-l} \sqrt{l(l+1)l'(l'+1)} \frac{j_l(k\eta_0)
   j_{l'}(k\eta_0)}{(k\eta_0)^2} Y_{lm}^*(\mathbf{n}) Y_{l'm'}
   (\mathbf{n'}) \, \mathbf{Y}_{lm}^{(1)}(\mathbf{\hat{k}}) \cdot
   \mathbf{Y}_{l'm'}^{(1)\,*} (\mathbf{\hat{k}}).
   \label{spectruma}
\end{eqnarray}
In this result, the term proportional to $\hat{k}_i \hat{k}_j$ in the
projector $P_{ij}$ in the magnetic field power spectrum in
Eq.~(\ref{spectrum}) vanishes because $\mathbf{\hat k} \cdot
\mathbf{Y}_{lm}^{(1)}=0$. The term proportional to the magnetic field
helicity spectrum $P_H(k)$ is also identically zero because
$[\mathbf{\hat k} \times
 \mathbf{Y}^{(1)}_{lm}(\mathbf{\hat k})]
\cdot \mathbf{Y}^{(1)\,*}_{l'm'}(\mathbf{\hat k}) =
\mathbf{Y}^{(0)}_{lm}(\mathbf{\hat k}) \cdot
\mathbf{Y}^{(1)\,*}_{l'm'}(\mathbf{\hat k})=0$.  This result is in
agreement with Ensslin and Vogt \cite{ensslin03} and Campanelli et
al.~\cite{campanelli04}, who conclude that Faraday rotation cannot be
used to reconstruct the helical part of the magnetic field
spectrum. The opposite conclusion of Ref.~\cite{pogosian02} is
erroneous. Physically, in real space, non-zero helicity of the
magnetic field only affects the off-diagonal elements in the
correlation matrix of the magnetic field components, while the Faraday
rotation is due only to the diagonal component corresponding to the
propagation direction; see \cite{ensslin03} for a more detailed
discussion.

Using the orthogonality of vector spherical harmonics,
\begin{equation}
\int d\Omega_{\mathbf{\hat k}}
\mathbf{Y}_{lm}^{(\lambda)\,*}(\mathbf{\hat{k}})\cdot
\mathbf{Y}_{l'm'}^{(\lambda^\prime)}(\mathbf{\hat{k}})
=\delta_{\lambda\lambda^\prime}\delta_{ll'} \delta_{mm'},
\label{ort}
\end{equation}
and the usual spherical harmonic summation formula
\begin{equation}
P_l({\bf n}\cdot{\bf n'}) = {4\pi\over 2l+1}\sum_{m=-l}^l
Y^*_{lm}({\bf n}) Y_{lm}({\bf n'}),
\end{equation}
Eq.~(\ref{spectruma}) simplifies to
\begin{equation}
   \left\langle R({\bf n}) R({\bf n^\prime})\right\rangle
   \simeq\frac{9}{128\pi^5q^2}
   \sum_{l} \frac{2l+1}{4\pi}l(l+1)P_l({\bf n \cdot n'})
   \int dk \, k^2 P_B(k) \left(\frac{j_l(k\eta_0)}{k\eta_0}\right)^2,
   \label{spectruma2}
\end{equation}
with corresponding multipole moments
\begin{equation}
   C_l^R \simeq \frac{9 l(l+1)}{(4\pi)^3q^2}
   \frac{B^2_\lambda}{\Gamma\left(n_B/2 +3/2 \right)}
   \left(\frac{\lambda}{\eta_0}\right)^{n_B+3} \int_0^{x_D}
   dx \, x^{n_B}j^2_l(x),
   \label{ClRR-sym-int}
\end{equation}
where $x_D=k_D\eta_0$ and the multipole moments are defined via
\begin{equation}
   \left\langle R({\bf n}) R({\bf n^\prime})\right\rangle
   = \sum_{l} \frac{2l+1}{4\pi} C_l^R P_l ({\bf n \cdot n^\prime}).
\end{equation}
The rotation angle power spectrum is simply given by the rescaling
\begin{equation}
   C_l^\alpha=\nu_0^{-4} C_l^R.
   \label{Cla}
\end{equation}
Note that Refs.\ \cite{pogosian02, harari97} have
incorrect prefactors in their corresponding expressions for the
rotation multipoles $C_l^R$, due to the non self-consistent choice
of the units.

This rotation multipole expression contains a sharp short-wavelength
cutoff $k_D$; in reality, the effective cutoff will be smoothly spread
over a range of scales.  To prevent unphysical oscillations in the
integral of Eq.~(\ref{ClRR-sym-int}), and to simplify the numerical
evaluation of the integral, we replace the oscillatory function
$j^2_l(x)$ by half of its envelope, $1/(2x^2)$, for all $x$ larger
than the second zero of $j_l(x)$. Then the tail of the integrand is
just a trivial power law, and the total integral will not show
oscillations, as would be expected from a more realistic cutoff
function.

\section{$C$-Polarization from Faraday Rotation}

The preceding Section gives an expression for the power spectrum of
an approximation to the polarization orientation Faraday rotation field
$\alpha({\bf n})$ which can be applied to the unrotated polarization
field of the microwave background radiation.
If the unrotated polarization field
arises only from scalar perturbations it will only have a nonzero
$G$ polarization component. Faraday rotation will induce non-zero $C$
polarization; here we compute the $C$-polarization power spectrum. We
use the differential geometry formalism of Ref.~\cite{kks97b} for this
computation, and work with tensors defined on the two-dimensional
spherical manifold representing the sky.

We represent the unrotated polarized CMB with the tensor field
$P_{ab}(\hat{\bf n})$ on the two-dimensional sphere orthogonal to
the direction vector ${\bf n}$, and decompose this in terms of tensor
spherical harmonics in the usual way
\begin{equation}
   P_{ab} = \sum_{lm}\left[ a^G_{lm} Y^G_{(lm)ab}
   + a^C_{lm} Y^C_{(lm)ab}\right] .
   \label{multipoles}
\end{equation}
Here the usual $lm$ indices are enclosed in parentheses to distinguish
them from the $ab$ tensor indices. Explicit forms for the orthonormal
tensor spherical harmonics are
\begin{eqnarray}
   Y^G_{(lm)ab}({\bf n}) &=& N_l\left(Y_{(lm):ab}({\bf n})
   + {1\over 2}g_{ab}l(l+1)Y_{(lm)}({\bf n})\right),\nonumber\\
   Y^C_{(lm)ab}({\bf n}) &=& {N_l\over 2}\left(Y_{(lm):ac}
   ({\bf n})\epsilon^c_b
   + Y_{(lm):bc}({\bf n})\epsilon^c_a\right) ,
   \label{tens_harmonics}
\end{eqnarray}
with the normalization factor $N_l = (2(l-2)!/(l+2)!)^{1/2}$. Here
$g_{ab}$ is the metric on the sphere, $\epsilon_{ab}$ is the
antisymmetric Levi-Civita tensor, and the covariant derivative on the
sphere is $A({\bf n})_{:a}\equiv\nabla_{\bf n}A({\bf n})$, using the
notation of the previous Section.
For polarization from scalar perturbations, $a^C_{lm}=0$
\cite{kamionkowski97a,seljak97}.
Primordial tensor perturbations, or the vector and tensor
perturbations induced by magnetic fields, will give
nonzero $a^C_{lm}$.

A rotation of the polarization orientation
by an angle $\alpha$ is represented by the transformation
\begin{equation}
   P^\prime_{ab} = R_a{}^c P_{cb}
\end{equation}
with the rotation operator given by
\begin{equation}
   R_a{}^c = \cos(2\alpha) g_a{}^c + \sin(2\alpha) \epsilon_a{}^c.
   \label{rotationmatrix}
\end{equation}
In the limit of small rotation angle, $\alpha\ll 1$, the rotation
operator reduces to $R_a{}^c\simeq g_a{}^c + 2\alpha\epsilon_a{}^c$,
which is linear in $\alpha$.
In our case the rotation field varies with
direction $\alpha=\alpha ({\bf n})$, so we decompose it into
spherical harmonics
\begin{equation}
   \alpha({\bf n}) = \sum_{lm} b_{lm}Y_{(lm)}({\bf n}).
\label{rotationfield}
\end{equation}

We now use Eq.~(\ref{multipoles}) to expand the resulting polarization
field in tensor spherical harmonics.
Inverting for the coefficients of the rotated $C$ polarization we have,
\begin{eqnarray}
   a_{lm}^{C}{}' &=& \int d\Omega_{\bf n}
   P'_{ab}({\bf n})Y_{(lm)}^{C\,ab\,*}({\bf n})
   = 2\int d\Omega_{\bf n} \alpha({\bf n}) \epsilon_a{}^cP_{cb}({\bf n})
   Y_{(lm)}^{C\,ab\,*}({\bf n}) \nonumber \\
   &=& 2\sum_{l_1m_1}\sum_{l_2m_2} b_{l_1m_1}a^G_{l_2m_2}
   \int d\Omega_{\bf n} Y_{(l_1m_1)}({\bf n})~\epsilon_a{}^c
   Y^G_{(l_2m_2)\,cb}({\bf n}) Y^{C\,ab\,*}_{(lm)}({\bf n}),
   \label{almcprime}
\end{eqnarray}
and likewise for the $G$-polarization multipole coeffecients,
\begin{equation}
   a_{lm}^G{}'
   = 2\sum_{l_1m_1}\sum_{l_2m_2} b_{l_1m_1}a^C_{l_2m_2}
   \int d\Omega_{\bf n} Y_{(l_1m_1)}({\bf n})~\epsilon_a{}^c
   Y^C_{(l_2m_2)\,cb}({\bf n}) Y^{G\,ab\,*}_{(lm)}({\bf n}) .
   \label{almgprime}
\end{equation}

The integrals over three spherical harmonics can be evaluated through
repeated integrations by parts, after making use of the form of the
tensor spherical harmonics and the identities
\begin{equation}
   \epsilon_{ac}\epsilon_{bd} = g_{ab}g_{cd} - g_{ad} g_{cb},\qquad
   \epsilon_{ca} \epsilon^{c}_{~b}=g_{ab}=-\epsilon_{ac}\epsilon_{~b}^c,\qquad
   \epsilon_{ab:c}=0
   \label{epsilon-identity}
\end{equation}
which eliminates the product of antisymmetric tensors in the integrand.
We also need the eigenvalue equation
$Y_{(lm):a}{}^{:a}({\bf n}) = -l(l+1)Y_{(lm)}({\bf n})$
and use the integration by parts formula
\begin{equation}
   \int d\Omega_{\bf n} A_{:a} B = -\int d\Omega_{\bf n} A B_{:a},
   \label{byparts}
\end{equation}
which has no surface term contribution as long as the integration
is over the entire sphere. Using the explicit forms for
$Y_{(lm)ab}^G({\bf n})$ and $Y_{(lm)ab}^C({\bf n})$ in
Eqs.~(\ref{tens_harmonics}) and using Eq.~(\ref{epsilon-identity})
results in
\begin{eqnarray}
   \int d\Omega_{\bf n} Y_{(l_1 m_1)}({\bf n})~\epsilon_a{}^c
   Y^G_{(l_2 m_2)cb}({\bf n})Y_{(l m)}^{C\,ab\,*}({\bf n}) &=&
   {N_{l_2}N_{l}\over 2}\int d\Omega_{\bf n} Y_{(l_1 m_1)}({\bf n})
   Y_{(l_2 m_2):b}{}^{:b}({\bf n})
   Y^*_{(l m):a}{}^{:a}({\bf n})\nonumber\\
   &{ }& - N_{l_2}N_{l}\int d\Omega_{\bf n} Y_{(l_1 m_1)}({\bf n})
   Y_{(l_2 m_2):ab}({\bf n})Y^*_{(lm)} {}^{:ab}({\bf n}).
   \label{simpler}
\end{eqnarray}
For $G\leftrightarrow C$ in the integrand the result is the
same except for an overall sign change.
The derivatives in the first integral just add $l(l+1)$ factors.

The second integral is more involved, and is computed in Appendix
A. Substituting Eqs.~(\ref{I1eval}) and (\ref{3Y}) into
Eq.~(\ref{simpler}) and then into Eq.~(\ref{almcprime}) gives finally
\begin{equation}
   a_{lm}^{C}{}' =N_l \sum_{l_1m_1}\sum_{l_2m_2}
   N_{l_2} K(l,l_1,l_2) b_{l_1m_1}a^G_{l_2m_2}
   \int d\Omega \, Y_{(l_1m_1)} Y_{(l_2m_2)} Y^*_{(lm)},
   \label{aC_int}
\end{equation}
where
\begin{equation}
   K(l,l_1,l_2)\equiv -{1\over 2}\left(L^2 + L_1^2 + L_2^2 -2L_1L_2
   -2L_1L +2L_1-2L_2 -2L\right),
\end{equation}
with $L=l(l+1)$, $L_1=l_1(l_1+1)$, and $L_2=l_2(l_2+1)$.
The integral over three spherical harmonics is a well-known expression
in terms of Clebsch-Gordan coefficients, Eq.~(\ref{3Y}).
The coefficient $a_{lm}^G{}'$ is given by the same formula with
the replacements $a^G_{l_2m_2}\rightarrow a^C_{l_2m_2}$ and
$N_l\rightarrow -N_l$.

To derive the power spectrum, we need to evaluate
\begin{equation}
   \langle a_{l'm'}^{C}{}'{}^*a_{lm}^C{}'\rangle \equiv
   C^C_l\delta_{ll'}\delta_{mm'} ,
   \label{asquare}
\end{equation}
where the definition of $C^C_l$ holds
for any statistically isotropic distribution; here we assume that
all relevant cosmological quantities satisfy this condition.
Assuming that the rotation field and the original temperature
polarization field are uncorrelated, we have
\begin{equation}
   \langle b_{l_1m_1}^*b_{l_2m_2}a_{l_3m_3}^{G\,*}a_{l_4m_4}^G\rangle
   =\delta_{l_1l_2}\delta_{m_1m_2}\delta_{l_3l_4}\delta_{m_3m_4}C_{l_1}^\alpha
   C_{l_3}^G .
   \label{fourpoint}
\end{equation}
where $C^\alpha_l$ is the angular power spectrum of the rotation field
and $C^G_l$ is the $G$-polarization power spectrum of the unrotated
polarization field just after decoupling. Using the expressions
given in Eqs.~(\ref{aC_int}) and (\ref{fourpoint}) in Eq.~(\ref{asquare})
results in a long expression with sums over $l_1m_1$ and $l_2m_2$. This
can be further simplified using the Clebsch-Gordan identity
\begin{equation}
   \sum_{m_1m_2} C^{lm}_{l_1m_1l_2m_2}C^{l'm'}_{l_1m_1l_2m_2}
   = \delta_{ll'}\delta_{mm'} ,
   \label{CCsum}
\end{equation}
and we find
\begin{equation}
   C^C_l= N_l^2 \sum_{l_1l_2}
   N_{l_2}^2 K(l,l_1,l_2)^2 C^G_{l_2}C^\alpha_{l_1}
   {(2l_1+1)(2l_2+1)\over 4\pi(2l+1)}\left(C^{l0}_{l_10l_20}\right)^2 .
   \label{answer}
\end{equation}
This is our basic result for the $C$-polarization power spectrum
induced by Faraday rotation of the primordial $G$-polarization; we
numerically evaluate it in the following Section. (Numerical techniques
for evaluating the remaining Clebsch-Gordan coefficients are given in
Appendix B.) Note that the more familiar multipole coefficients $C^E_l$
and $C^B_l$ \cite{zaldarriaga97} are just a factor of two larger than
$C^G_l$ and $C^C_l$, respectively.

We note that, for a stochastic magnetic field, the
cross-correlations between the intrinsic temperature and
$G$-polarization fluctuations from non-magnetic perturbations and the
$C$-polarization from Faraday rotation result in $C_l^{TC}=0$. These
multipoles are linear in the magnetic field so the ensemble average
vanishes. Non-zero correlations result only from the intrinsic
$T$ fluctuations arising from the magnetic field
itself and the $C$-polarization from Faraday rotation, but this
case is not considered here. The $C_l^{GC}$ correlation will also be
zero unless the magnetic field has non-zero helicity,
which follows from the invariance of
$\left\langle QU\right\rangle$ under Faraday rotation by a
stochastic magnetic field \cite{mel98}\footnote{non-zero off diagonal
$C_l^{GC}$ correlation will apppear in the case of homogeneous magnetic field
Ref.~\cite{scoccola04}}.

\section{Numerical Results}

\begin{figure*}
\includegraphics[angle=270, width=6.5in]{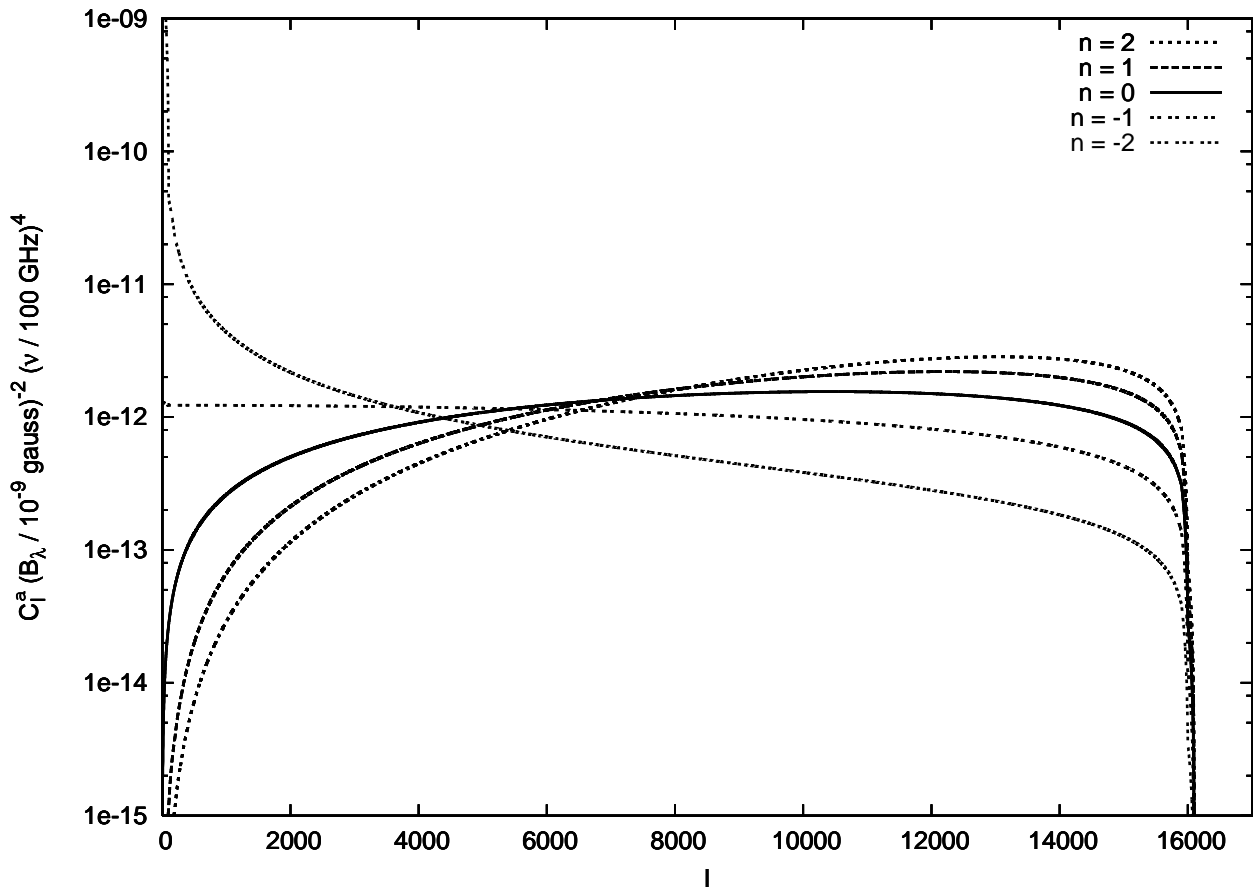}
\caption{The angular power spectrum of the Faraday rotation angle,
$C_l^\alpha$, induced by a stochastic magnetic field. The curves in
order of decreasing amplitude on the right side of the plot correspond
to magnetic field power spectral indices $n_B=2$, 1, 0, $-1$, and $-2$.
The magnetic fields have been normalized to a nanogauss at the smoothing
scale $\lambda = 1$ Mpc.}
\label{anglefig}
\end{figure*}

The rotation angle power spectrum, Eq.~(\ref{Cla}), is shown in Fig.\
\ref{anglefig} for values of the magnetic field spectral index ranging
from $n_B=-2$ to $n_B=2$. The plotted rotation power spectra are for a
magnetic field with $B_\lambda=10^{-9}$ G and an observation frequency
of $\nu=100$ GHz; the rotation power scales like
$B_\lambda^2\nu^{-4}$. We have assumed a cutoff scale $k_D = 2.0$
Mpc$^{-1}$ approximately corresponding to the Silk damping scale. The
polarization has a root-mean-square rotation angle given by
\begin{equation}
   \bar\alpha\equiv
   \left\langle\alpha^2\right\rangle^{1/2} =\left[\sum_l {2l+1\over 4\pi}
   C_l^\alpha\right]^{1/2}.
   \label{rmsalpha}
\end{equation}
At $\nu=100$ GHz and $B_\lambda=10^{-9}$ G with $\lambda = 1$ Mpc,
${\bar\alpha}\approx 0.3^\circ$ for all considered values of
$n_B$. This is consistent with $\bar\alpha$ found in
Ref.~\cite{kosowsky96} for a constant magnetic field of strength
$B_\lambda$ (around $1.6^\circ$ at $\nu=30$ GHz). For negative values
of $n_B+1$, the magnetic field power spectrum grows with length scale, so
the largest rotations are seen at large scales corresponding to small
$l$ values. The opposite is true for positive values of $n_B+1$, which
increase in power at smaller scales and larger $l$ values.

\begin{figure*}
\includegraphics[angle=270, width=6.5in]{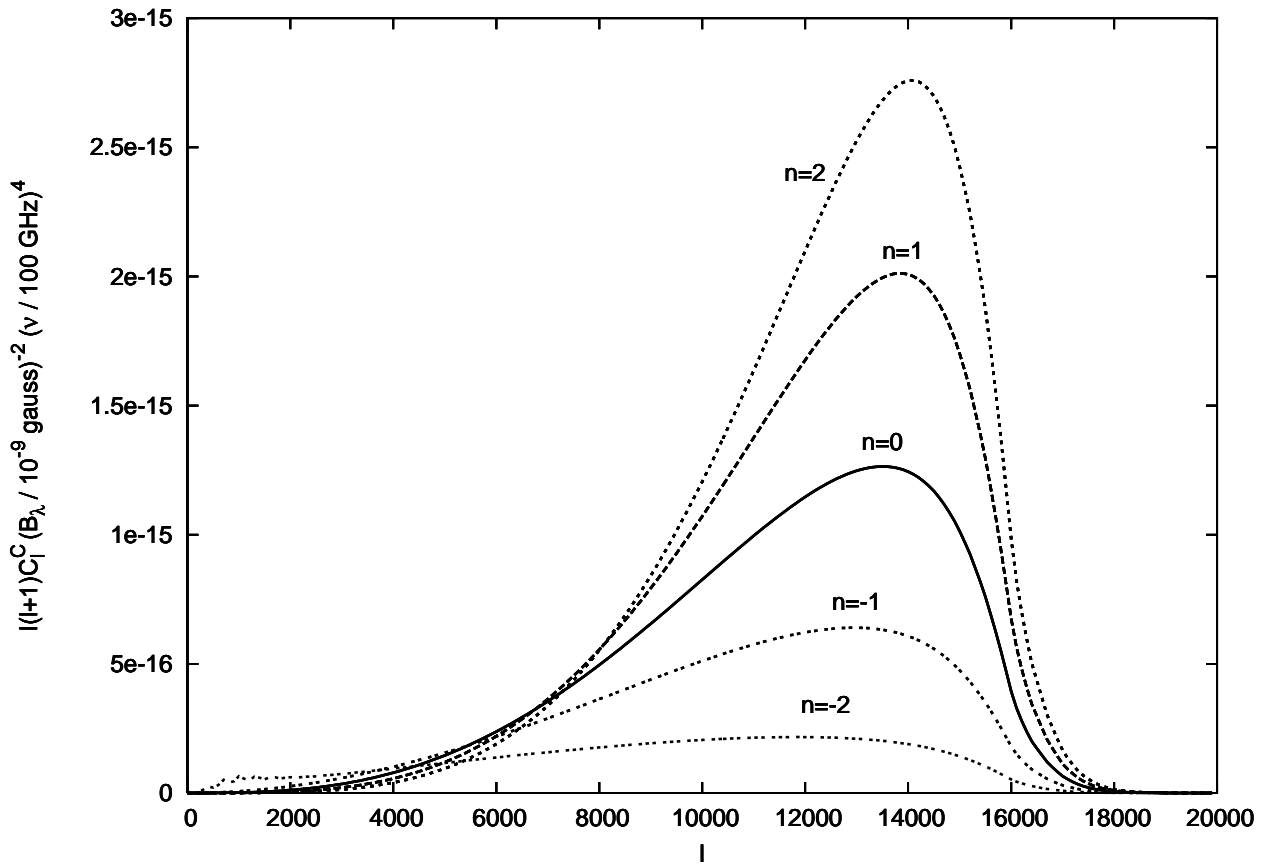}
\caption{The $C$-polarization power spectrum of the microwave background
induced by the Faraday rotation field in Fig.~\ref{anglefig}, again
with the magnetic field normalization scale $\lambda = 1$ Mpc.}
\label{cpowerfig}
\end{figure*}

The power spectrum of $C$-polarization induced by this Faraday rotation
power spectrum is displayed in Fig.~\ref{cpowerfig}. This power
spectrum is also for $B_\lambda=10^{-9}$ G and $\nu=100$ GHz, scaling
like $B_\lambda^2\nu^{-4}$. The most notable feature of the
$C$-polarization spectrum is its peak at small angular scales at $l>10^4$.
The cutoff at $l=16000$ is sharper than in a more realistic damping model
where it occurs over a range of scales somewhat past $k_D$, but the peak
position and amplitude would be largely unchanged by a different
damping prescription.  We also assume a standard $\Lambda$CDM
cosmological model and compute the $C_l^G$ power spectrum appearing in
Eq.~(\ref{answer}) to $l=5000$ using CMBFAST \cite{cmbfast}.

The peak angular scale is readily understood: at the last scattering
surface Faraday rotation induces polarization fluctuations on angular
scales corresponding to the characteristic wavelengths of the
stochastic magnetic field; just subsequent to recombination the
polarization is imprinted with fluctuations on scales much smaller
than the quadrupole scale which induced the polarization fluctuations
in the first place. Free streaming then shifts these fluctuations to
even smaller scales. So the Faraday rotation
power spectrum provides a distinctive signature of primordial magnetic
fields.

This computation only applies to small rotation angles, which for a
given magnetic field translates into a limiting lower observation
frequency. The root-mean-square rotation $\bar\alpha$,
Eq.~(\ref{rmsalpha}), scales like $B_\lambda \nu_0^{-2}$. For a field
strength of $10^{-9}$ G, observations at $\nu=30$ GHz yield a mean
rotation of around $3^\circ$, well within the small rotation limit,
but observations at $\nu=10$ GHz would give mean rotation angles of
around $30^\circ$, for which the small-angle approximation employed in
Eq.~(\ref{almcprime}) clearly is not valid.  If the rotation angle becomes
large, depolarization effects become important and the mean
polarization amplitude is reduced \cite{harari97}.

The other sources of cosmological $C$-polarization have been widely
discussed. The first is tensor (or vector) primordial
perturbations. In particular, an epoch of inflation in the early
universe necessarily produces nearly scale-invariant tensor as well as
scalar perturbations, and for many inflation models these tensor modes
are observable through the $C$-polarization they produce
\cite{kamionkowski98}. This class of tensor perturbation signals all
have the property that their $C$-polarization power peaks at large
scales, $l\lesssim 100$. The other generic source of $C$ polarization
is gravitational lensing of the primordial $G$ polarization from
scalar perturbations \cite{zaldarriaga98}, which peaks at angular
scales in the region of $l\approx 1000$. The $C$ polarization from
Faraday rotation peaks at significantly smaller angular scales, well
separated from either the tensor or lensing signals, and would thus be
easy to separate.

\section{Discussion and Conclusions}

Primordial magnetic fields produce $C$-polarization fluctuations
directly via the vector and tensor perturbations they induce in the
primordial plasma \cite{mack02}. For a field with $B_\lambda = 10^{-9}$
G, the peak polarization amplitude is, for example,
$l^2C^C_l\approx 10^{-13}$ at $l=1000$ for $n=-2$ \cite{lewis04}. The
Faraday rotation signal from the same magnetic field gives a peak
polarization amplitude of around $10^{-14}$ for $\nu=30$ GHz and
(neglecting the depolarization effect at significant rotation angles)
$10^{-12}$ for $\nu=10$ GHz. The direct polarization fluctuations peak
at significantly larger scales than those from Faraday rotation:
primordial magnetic fields have a double signature, with two
polarization peaks.  In comparison, the $G$ polarization power
spectrum from primordial density perturbations peaks at a comparable
amplitude of around $l^2C^G_l\approx 3\times 10^{-12}$ at an angular
scale of around $l=1000$, while the $C$ polarization from
gravitational lensing has an amplitude on the order of $l^2C^C_l
\approx 10^{-14}$ at $l=1000$. The $C$-polarization from Faraday
rotation for realistic primordial magnetic fields is small but
potentially measurable, in the same ballpark as detecting the lensing
polarization signal.

Several claims have been made that primordial magnetic fields of the
amplitudes considered here are ruled out. These claims are not all
compelling; we review some here. Caprini and Durrer have asserted
remarkably stringent limits on primordial magnetic fields through
their conversion into gravitational waves, which then are constrained
by the expansion rate of the universe at nucleosynthesis
\cite{caprini02}. But the expansion rate of the universe is the same
whether energy density is converted from magnetic fields to
gravitational waves or not, since the energy density of both scale
the same way with the expansion of the universe. So the actual
constraint is on the total radiation energy density in the magnetic
field, which is constrained to be about 1\% of the total energy
density in the usual manner that nucleosynthesis
limits extra neutrino species. The corresponding limit on the
total comoving mean magnetic field strength is around $10^{-8}$ gauss,
not the $10^{-27}$ gauss claimed in \cite{caprini02}.

Potentially more interesting limits come from magnetic fields in
galaxy clusters. It is believed that clusters contain magnetic
fields with $\mu$G field strengths, from Faraday rotation of
background polarized radio sources observed through the cluster
(e.g., \cite{taylor93,eilek02}). Vogt and Ensslin \cite{Vogt03}
have recently analyzed several cluster rotation maps, concluding
that fields of at least a few $\mu$G are present, with a lower
limit on the steepness of the magnetic field power spectrum in the
clusters. On the other hand, simulations starting from constant
comoving magnetic fields of $10^{-11}$ G show clusters generating
fields sufficiently large to explain Faraday rotation measurements
\cite{dolag99,dolag02}. Banerjee and Jedamzik \cite{banerjee04}
have claimed that these results rule out primordial fields with
comoving amplitudes as large as $10^{-9}$ G, arguing that they
would overproduce fields in galaxy clusters today. We feel this is
a potentially powerful argument, but premature. First, simulations
have shown that primordial fields which are 1\% of those
considered in this paper may lead to observed cluster fields, but
magnetic field generation is a non-linear process and it is not
obvious that the final cluster fields will scale linearly with the
initial fields; we know they will definitely not as equipartition
field strength is approached in the cluster. While it seems likely
that cluster fields are generally below equipartition value
\cite{clarke01}, simulations over a wide range of initial field
strengths and configurations must be performed to rule out certain
initial field configurations. Second, we do not fully understand
galaxy clusters, and it is not immediately clear whether the
cluster simulations to date include all of the relevant physics.
Third, the Faraday rotation signal in galaxy clusters depends on
not only the field strength but also its configuration: it is
conceivable that larger initial fields might lead to larger
cluster fields, but that much of the field strength would be
concentrated at small scales due to more efficient turbulent
cascades, and small-scale fields contribute much less to observed
Faraday rotation measurements.  It is likely that further cluster
simulations will resolve these issues; until then it is important
to examine other techniques for probing magnetic fields like the
one outlined in this paper.

Another constraint on primordial fields comes from field dissipation
on small scales \cite{jedamzik00} inducing spectral distortions in
the microwave background radiation. Field strengths greater than
$3\times 10^{-8}$ G on scales of 400 pc are ruled out. This limit
constrains primordial magnetic fields which are blue, $n>0$, with
most field strength on small scales. For the fields considered in
this paper, on scales larger than 1 Mpc, such constraints are
important if the magnetic field's power spectrum remains a power law
over a range of $10^4$ in length scale. As this may not be the case,
we present results in this paper for various values of $n$, but the
reader should note that pure power-law fields with $n>0.5$ must have
amplitudes smaller than $10^{-9}$ G on Mpc scales.

A signal we have not computed here explicitly, but which follows
simply from the formulas derived above, is the $G$-polarization signal
arising from the Faraday rotation of a primordial $C$-polarization
signal. Of course, scalar perturbations provide no such primordial
signal, but any magnetic field does \cite{mack02}. It is likely that
the magnetic field polarization signal is small compared to the
$G$-polarization from primordial density perturbations, but the
rotated $G$ polarization appears, like the $C$ polarization, at small
angular scales of $l\approx 10000$. At these small scales, any direct
primordial signals will be suppressed due to diffusion damping, so
this $G$ polarization signal would provide confirmation of the
corresponding $C$ polarization rotation signal in
Fig.~\ref{cpowerfig}.

Additionally, comparison of microwave background imprints from Faraday
rotation and intrinsic fluctuations can probe the helicity of the
primordial field. Here we have demonstrated explicitly that any
helical part of the magnetic field gives zero Faraday rotation.
Helical stochastic magnetic fields do, however, produce non-zero
temperature and polarization fluctuations directly
\cite{pogosian02,caprini03}. By comparing the microwave background
power spectra at scales of $20^\prime$ with the polarization power
spectra at scales of $1^\prime$, the helical power spectrum $P_H(k)$
can be probed separately from the non-helical power spectrum $P_B(k)$.

In this paper, we have modelled the Faraday rotation signal by
separating the polarization generation and rotation processes through
a simple approximation. A more accurate calculation demands a full
numerical evolution of the coupled Boltzmann hierarchy of equations
describing photons, electrons, and magnetic and gravitational
fields. Such a code has been developed for the direct temperature and
polarization fluctuations from magnetic fields \cite{lewis04}, and at
first glance it would appear to be a simple matter to modify such a
code to include Faraday rotation, including a more accurate treatment
than presented here of combined polarization generation and rotation,
and damping at small scales or large magnetic field strengths.
However, currently used microwave background Boltzmann codes use
polarization basis functions corresponding directly to $G$ and $C$
polarization modes. While this provides formal simplicity and
conceptual clarity, Faraday rotation, described by the rotation
operator Eq.~(\ref{rotationmatrix}), does {\it not} correspond to a
simple rotation of $G$ polarization into $C$ polarization, but is
naturally expressed as a rotation mixing the $Q$ and $U$ Stokes
parameters. A straightforward numerical implementation of Faraday
rotation requires using a code employing the Stokes parameters as
variables. Techniques for doing this and reconstructing the $G$ and
$C$ power spectra are known (see, e.g., \cite{kosowsky96a}) and were
used in a number of older microwave background polarization codes.

The polarization signals discussed in this paper have an amplitude of
$\mu$K or smaller, on angular scales of $1'$ or smaller. The
combination of small amplitude and angular scale make detection of
this potential signal challenging. For a given magnetic field
strength, the Faraday rotation power spectrum scales like $\nu^{-4}$
with observing frequency, so going to a low enough frequency will
compensate for smaller magnetic field strengths. But lower-frequency
observations at the same angular resolution require larger
experiments, and to obtain arcminute resolution at frequencies of 30
GHz or below over a reasonable field of view likely requires
interferometric experiments (the corresponding diffraction limit would
require a single-dish diameter of 50 meters). No polarized
interferometers acting at these frequencies and angular scales are
currently envisioned, but polarization experiments have only just
begun to detect any signals at all \cite{DASI,WMAPpol,CBI,CAPMAP}, and
we can expect rapid advances over the coming decade.  The new
generation of millimeter-wave frequency high-resolution experiments
for temperature fluctuations (e.g.,\ \cite{ACT}) will help lay the
technical basis for complementary polarization observations.

The Faraday rotation signal in Fig.~\ref{cpowerfig} is a demanding
experimental target, and its amplitude is highly uncertain (and could
certainly be zero). But we will not be obliged to go directly from
ignorance to probes of these small scales, because cosmological
magnetic fields strong enough to induce detectable Faraday rotation
will also cause temperature and polarization fluctuations at larger
scales. If primordial magnetic fields are present, we will suspect
their existence long before the Faraday rotation power spectrum is
probed directly, and the existence of detectable Faraday rotation can
be verified through multi-frequency polarization observations along
individual lines of sight. If the coming five years strengthen current
suspicions of a possible primordial magnetic field into more solid
evidence, then the Faraday rotation signal calculated in this paper
will become a compelling experimental opportunity to characterize a
new and poorly understood relic of the early universe.

\acknowledgments
We thank Ruth Durrer, Grigol Gogoberidze, Levon Pogosian, and Tanmay
Vachaspati for helpful discussions. TK and BR acknowledge support
from NSF CAREER grant AST-9875031 and DOE EPSCoR grant DE-FG02-00ER45824.
TK, AK, and GL are partially supported by the U.S.--Georgia Bilateral Program
of the Civilian Research and Development Foundation, grant 3316. AK is a
Cottrell Scholar of the Research Corporation.

\begin{appendix}

\section{Integrals Involving Three Tensor Spherical Harmonics}

The evaluation of the needed integrals over three tensor spherical
harmonics can be built on the integral over three usual (scalar)
spherical harmonics using repeated integration by parts.
The integral over three spherical harmonics is well known
\cite{varshalovich89}
,
\begin{equation}
   \int d\Omega \, Y_{(l_1m_1)} Y_{(l_2m_2)} Y^*_{(lm)}=
   \left((2l_1+1)(2l_2+1)\over 4\pi(2l+1)\right)^{1/2}
   C^{l0}_{l_10l_20} C^{lm}_{l_1m_1l_2m_2},
\label{3Y}
\end{equation}
where $C^{lm}_{l_1m_1l_2m_2}$ is a Clebsch-Gordan coefficient.

Now we need to compute the second integral in Eq.~(\ref{simpler}),
\begin{equation}
   I_1\equiv\int d\Omega_{\bf n} Y_{(l_1 m_1)}({\bf n})
   Y_{(l_2 m_2):ab}({\bf n})Y^*_{(l m)} {}^{:ab}({\bf n}).
\label{Idef}
\end{equation}
First consider the simpler integral
\begin{equation}
   I_2\equiv\int d\Omega_{\bf n} Y_{(l_1 m_1)}({\bf n}) Y_{(l_2 m_2):a}
   ({\bf n})
 Y^*_{(l_3 m_3)}{}^{:a}({\bf n}).
   \label{a11}
\end{equation}
This can be evaluated by integrating by parts three times: first
on the derivative of $Y_{(l_3 m_3)}$, which gives two terms, for
one of which the derivatives change into $l_2(l_2+1)$. Now for the
other term, integrate by parts a second time on the derivative of
$Y_{(l_2 m_2)}$, and then follow by a third integration by
parts on the derivative of $Y_{(l_1 m_1)}$. The integral which is
left is the negative of the original integral; solving for this
integral results in
\begin{equation}
   I_2 = {1\over 2} \left(-L_1 + L_2 + L_3\right)
   \int d\Omega_{\bf n} Y_{(l_1 m_1)}({\bf n}) Y_{(l_2 m_2)}({\bf n})
   Y_{(l_3 m_3)}^*({\bf n})
   \label{I2eval}
\end{equation}
with the abbreviation $L_i\equiv l_i(l_i+1)$. The remaining
integral is just Eq.~(\ref{3Y}).

Using Eq.~(\ref{I2eval}), the integral (\ref{simpler}) can be
evaluated using the same general strategy, with three
integrations by parts. In this case, we also need to compute
$Y_{:ab}{}^{:b} = Y_{:ba}{}^{:b}$. To simplify this, we need to commute
the two outer derivatives. The curvature tensor for this 2-dimensional
maximally symmetric space is just
\begin{equation}
   R_{abcd} = g_{ac} g_{bd} - g_{ad}g_{bc}.
\end{equation}
Thus we have
\begin{equation}
   Y_{:ab}{}^{:b}=Y_{:ba}{}^{:b}=Y^{:b}{}_{:ba}+R_{dba}{}^b Y^{:d}
   =-[l(l+1)-1]Y_{:a}.
\label{Yabb}
\end{equation}
Now we have all the pieces needed to evaluate
Eq.~(\ref{Idef}): after three integrations by parts, we are
again left with the original integral, plus surface terms,
like in the case of Eq.~(\ref{a11}). The surface terms can
all be evaluated in closed form with the help of Eqs.~(\ref{Yabb}) and
(\ref{I2eval}), giving
\begin{equation}
   I_1 = {1\over 4}(L_2+L-L_1 - 2)(L_2+L-L_1)\int d\Omega_{\bf n}
   Y_{(l_1 m_1)}Y_{(l_2 m_2)}Y^*_{(lm)}.
   \label{I1eval}
\end{equation}
Again, the remaining integral is given by Eq.~(\ref{3Y}).

\section{Clebsch-Gordan Coefficient Numerics}

The Clebsch-Gordan coefficients in Eq.~(\ref{answer})
can be evaluated in closed form as \cite{varshalovich89}
\begin{equation}
   C^{c0}_{a0b0} = {(-1)^{g-c}\sqrt{2c+1}g!\over(g-a)!(g-b)!(g-c)!}
   \left[(2g-2a)!(2g-2b)!(2g-2c)!\over(2g+1)!\right]^{1/2},
   \label{Cc0a0b0}
\end{equation}
which is valid for $a+b+c=2g$ with $g$ a positive integer; $a$, $b$,
and $c$ must also satisfy the triangle inequalities. If
$a+b+c$ is odd or if $a$, $b$, and $c$ cannot form the sides of
a (possibly degenerate) triangle, the coefficient vanishes.
This coefficient can
be directly evaluated numerically, as long as the various
factors are taken in an order to prevent overflow and/or underflow errors.
However, direct evaluation is not fast enough for efficient evaluation
of Eq.~(\ref{answer}), since performing the sums up to $l$ values
of several thousand demands millions of Clebsch-Gordan coefficients,
most with very large indices. But the two-term Stirling asymptotic
expansion
\begin{equation}
   n!\sim n^ne^{-n}\sqrt{2\pi n}\left(1+{1\over 12n}\right),
   \qquad n\rightarrow\infty,
   \label{stirling}
\end{equation}
is accurate to 0.1\% even for $n=1$. Applying this expansion to all
of the factorial factors in Eq.~(\ref{Cc0a0b0}) gives the approximation
\begin{equation}
   {\left(C^{c0}_{a0b0}\right)^2\over 2c+1} \approx {e\over 2\pi}
   \left(1+{1\over 2g}\right)^{-2g-3/2}
   \exp\left({1\over 8g}-{1\over 8(g-a)}-{1\over 8(g-b)} -
   {1\over 8(g-c)}\right)
   \left[g(g-a)(g-b)(g-c)\right]^{-1/2},
   \label{C_asymp}
\end{equation}
when $g-a$, $g-b$, and $g-c$ are all non-zero.
This approximation is accurate to just over 1\% for the case $a=b=c=2$,
and better than 1\% for all other needed terms. For the degenerate
case $a+b=c$,
\begin{equation}
   {\left(C^{c0}_{a0b0}\right)^2\over 2c+1} \approx {e\over 2\sqrt{\pi}}
   \left(1+{1\over 2g}\right)^{-2g-3/2}
   \exp\left({1\over 8g}-{1\over 8a}-{1\over 8b}\right)
   \left(gab\right)^{-1/2},
   \label{C_asymp_degen}
\end{equation}
along with cyclic permutations for the other two cases; this approximation
is accurate to better than a part in $10^{-4}$.

\end{appendix}

\end{document}